\documentclass[preprint]{aastex}
\usepackage{graphicx}
\begin{document}

\newcommand{\um}{$\mu$m}

\newcommand\tna{\,\tablenotemark{a}}
\newcommand\tnb{\,\tablenotemark{b}}

\title{The Mid-Infrared Instrument for the James Webb Space Telescope, IV: The Low Resolution Spectrometer}
\author{Sarah~Kendrew\altaffilmark{1,2,3}, Silvia~Scheithauer\altaffilmark{2}, Patrice~Bouchet\altaffilmark{4},  Jerome~Amiaux\altaffilmark{4},    Ruym\'an~Azzollini\altaffilmark{5,6},   Jeroen~Bouwman\altaffilmark{2},   C. H.~Chen\altaffilmark{7}, D. Dubreuil\altaffilmark{4},  Sebastian~Fischer\altaffilmark{8,9},    Alistair~Glasse\altaffilmark{10},     T. P.  Greene\altaffilmark{11}, P.-O.~Lagage\altaffilmark{4},  Fred~Lahuis\altaffilmark{3,12},  Samuel~Ronayette\altaffilmark{4},   David~Wright\altaffilmark{13}, G. S.~Wright\altaffilmark{10}}

\altaffiltext{1}{University of Oxford, Department of Physics, Denys Wilkinson Building, Keble Road, Oxford, OX1 3RH, UK} 
\altaffiltext{2}{Max Planck Institute for Astronomy (MPIA), K\"onigstuhl 17, D-69117 Heidelberg, Germany}
\altaffiltext{3}{Leiden Observatory, Leiden University, PO Box 9513, 2300 RA, Leiden, The Netherlands.}
\email{sarah.kendrew@astro.ox.ac.uk}
\altaffiltext{4}{Laboratoire AIM Paris-Saclay, CEA-IRFU/SAp, CNRS, Universit\'e Paris Diderot, F-91191 Gif-sur-Yvette, France}
\altaffiltext{5}{Dublin Institute for Advanced Studies, School of Cosmic Physics, 31 Fitzwilliam Place, Dublin 2, Ireland}
\altaffiltext{6}{Centro de Astrobiolog\'ia (INTA-CSIC), Dpto Astrof\'isica, Carretera de Ajalvir, km 4,28850 Torrej\'on de Ardóz, Madrid, Spain}
\altaffiltext{7}{Space Telescope Science Institute, 3700 San Martin  Drive, Baltimore, MD, 21218, USA}
\altaffiltext{8}{Deutsches Zentrum f\"ur Luft- und Raumfahrt (DLR), K\"onigswinterer Str. 522-524, 53227, Bonn, Germany}
\altaffiltext{9}{I. Physikalisches Institut, Universit\"at zu K\"oln, Z\"ulpicher Str. 77,  50937 K\"oln, Germany}
\altaffiltext{10}{UK Astronomy Technology Centre, Royal Observatory,  Blackford Hill Edinburgh, EH9 3HJ, Scotland, United Kingdom}
\altaffiltext{11}{Ames Research Center, M.S. 245-6, Moffett Field, CA 94035, USA}
\altaffiltext{12}{SRON - Groningen Landleven 12 9747AD, The Netherlands}
\altaffiltext{13}{Stinger Ghaffarian Technologies, Inc., Greenbelt, MD, USA.}
\altaffiltext{}{}
\altaffiltext{}{}
\altaffiltext{}{}
\altaffiltext{}{}
\altaffiltext{}{}
\altaffiltext{}{}

\begin{abstract}
The Low Resolution Spectrometer of the MIRI, which forms part of the imager module, will provide R$\sim$100 long-slit and slitless spectroscopy from 5 to 12~\um. The design is optimised for observations of compact sources, such as exoplanet host stars. We provide here an overview of the design of the LRS, and its performance as measured during extensive test campaigns, examining in particular the delivered image quality, dispersion, and resolving power, as well as spectrophotometric performance, flatfield accuracy and the effects of fringing. We describe the operational concept of the slitless mode, which is optimally suited to transit spectroscopy of exoplanet atmospheres. The LRS mode of the MIRI was found to perform consistently with its requirements and goals. 
\end{abstract}
  
\keywords{JWST, MIRI, Low Resolution Spectroscopy (LRS), mid-infrared, spectroscopy}

\section{Introduction}\label{sec:intro}

The Low Resolution Spectrometer (LRS) forms part of the MIRI Imager module. By design, it delivers R$\sim$100 spectroscopy from 5 to 10~$\mu$m, optimised to provide single-source broadband spectra of compact sources; in practice the throughput is useful to 12~\um. The spectrometer can be operated in slit or slitless mode, with the spectrum dispersed by a double prism assembly (DPA) mounted in the imager filter wheel. The dispersing element is of the type originally proposed by Amici \citep{donati1862} and previously utilized in near-infrared astronomy \citep{baffa2001} but for the MIRI fabricated in ZnS and germanium \citep{fischer2008, rossi2008} to allow operation in the mid-IR. The slit mask is mounted in a fixed position, allowing the spectrum to be dispersed in a fixed location on the Imager focal plane array. In slitless mode, the source can be placed anywhere on the array; a dedicated location and corresponding detector subarray were defined in testing to optimise flight operations. 

In the following sections, we will first describe the opto-mechanical design and build of the LRS module in the MIRI (\S2)  and a comprehensive overview of its measured performance based on extensive ground testing (\S3). In Section~\ref{sec:ops} we describe a number of operational procedures with the LRS, in particular the slitless spectroscopy mode. Throughout this paper we focus on LRS-specific measurements, procedures and results; we refer to accompanying publications in this volume for overviews of the MIRI Imager module, of which the LRS forms part~\citep[][Paper III]{paper3}; the MIRI focal plane system~\citep[][Paper VIII]{paper7a}; and the instrument-level operational concept~\citep[][Paper X]{paper9}. Key science with LRS is described by~\citet[][Paper I]{rieke2014}. 

\section{Opto-mechanical design description}\label{sec:design}

\begin{figure}
\centering
\includegraphics[width=5.0in.]{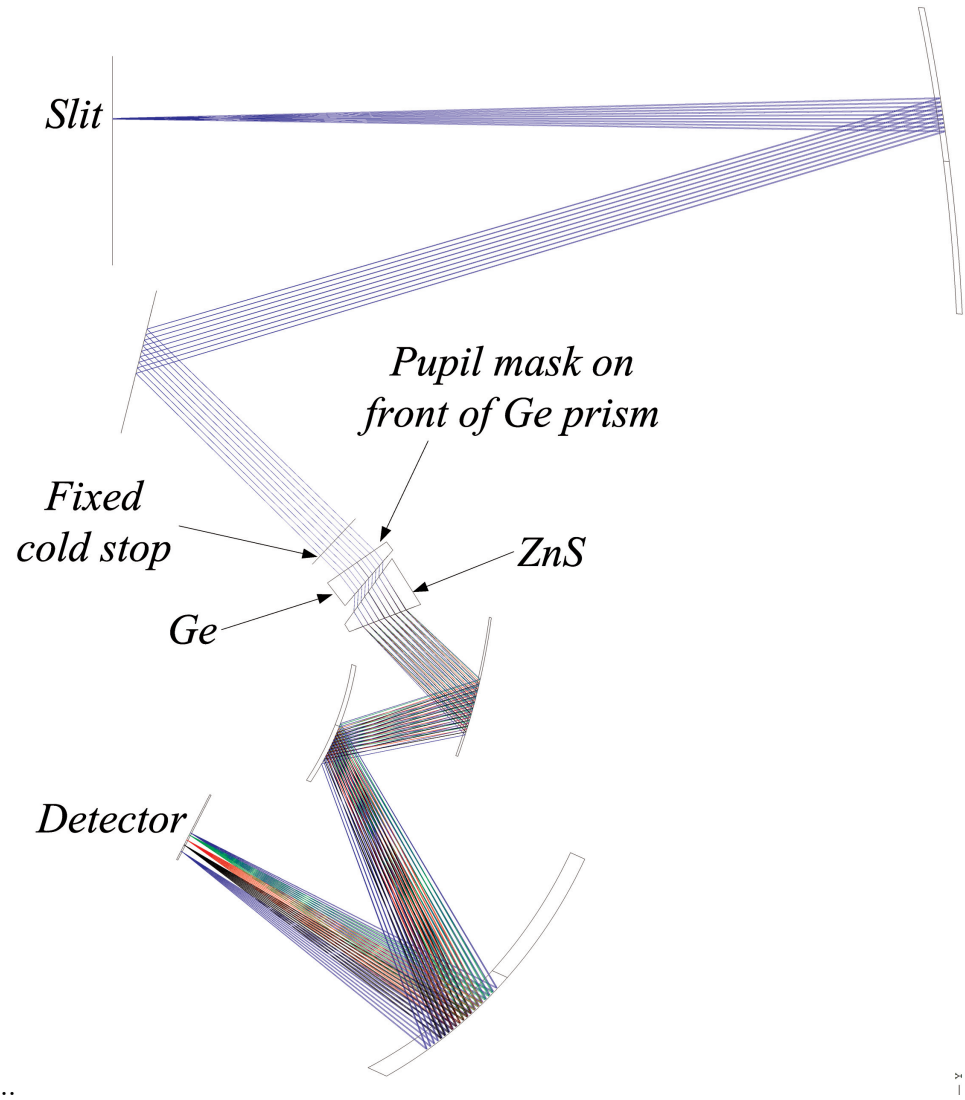}
\caption{Optical layout of the MIRIM/LRS module.}
\label{fig:lrs_optics}
\end{figure}

The LRS forms part of the MIRI imager module (MIRIM), which is described in detail in Paper III. As shown there, the MIRIM field of view is shared between the imager and coronagraph/LRS modes. The optical layout of the LRS is shown in Figure~\ref{fig:lrs_optics}; it is a collimator followed by a camera and shares all of its optics except the disperser with the imager functions of the MIRIM. The design includes a field stop, spectrometer slit and coronagraph masks at the telescope focal plane. After passing through the focal plane, the light is collimated and, at the pupil image formed by the collimator, a filter wheel holds the imaging filters, Lyot stops and filters for the coronagraph, and a pair of prisms for the LRS. A camera then images the field (or spectrum) onto the 1k $\times$ 1k detector at a final f-ratio of $\sim$f/7.14 giving a pixel scale of 0.11\arcsec/pixel. The MIRIM focal plane layout is shown in Figure~\ref{Fig:detector_with_spectrum}.

The LRS was designed to provide slit spectroscopy of compact sources from 5 to 10~\um~using a double prism mounted in the filter wheel. This produces a spectrum that covers some 140 detector pixels (3.483 mm), with a spectral resolving power varying approximately linearly from R = 40 at 5~\um~up to R = 160 at 10~\um.  The entrance slit is 4\farcs7~long and 0\farcs51~wide when projected onto the sky, corresponding to a length of 3.18~mm and a width of 0.33~mm in the focal plane, or 43 pixels and 4.6 pixels respectively, on the array (the LRS can also be used in a slitless mode). The as-designed spectral dispersion and resolving power are shown in Figure~\ref{fig:disp_res_design}. To mitigate the effect of the fold-over in the dispersion profile, a mask is mounted on the slit to block light at wavelengths short of 4.5~$\mu$m. The throughput is adequate to
provide a useful capability out to about 12-13 $\mu$m, particularly when operated in slitless mode, which is unaffected by transmission losses in the slit and slit mask (see Section~\ref{sec:slitvslitless}).

The spectroscopic dispersion is achieved with a pair of Ge/ZnS prisms, mounted at the filter wheel. The Ge prism cancels the deviation due to the ZnS dispersive prism. The mean thickness of the prisms is 7 mm and 11.06 mm for the Ge and the ZnS prism, respectively. Anti-reflective coatings for the prisms are optimized for the range 5~to 10~\um, the LRS nominal range.  The transmission is $>$90\% between 7~and 12~\um~for ZnS and $>$88\% between 7~and 13~\um~for Ge, and neither material shows any absorption in the range 5-10~\um. Therefore the expected efficiency for the set of prisms is near 80\% from $5 - 10 \mu$m (at cold temperatures) but drops below 25\% for wavelengths longer than 12 $\mu$m (if slit losses are included). The diaphragms associated with the prisms are without central obscuration and oversized, to take into account the telescope pupil shear and to avoid vignetting.

\begin{figure}[h]
\includegraphics[width=7in]{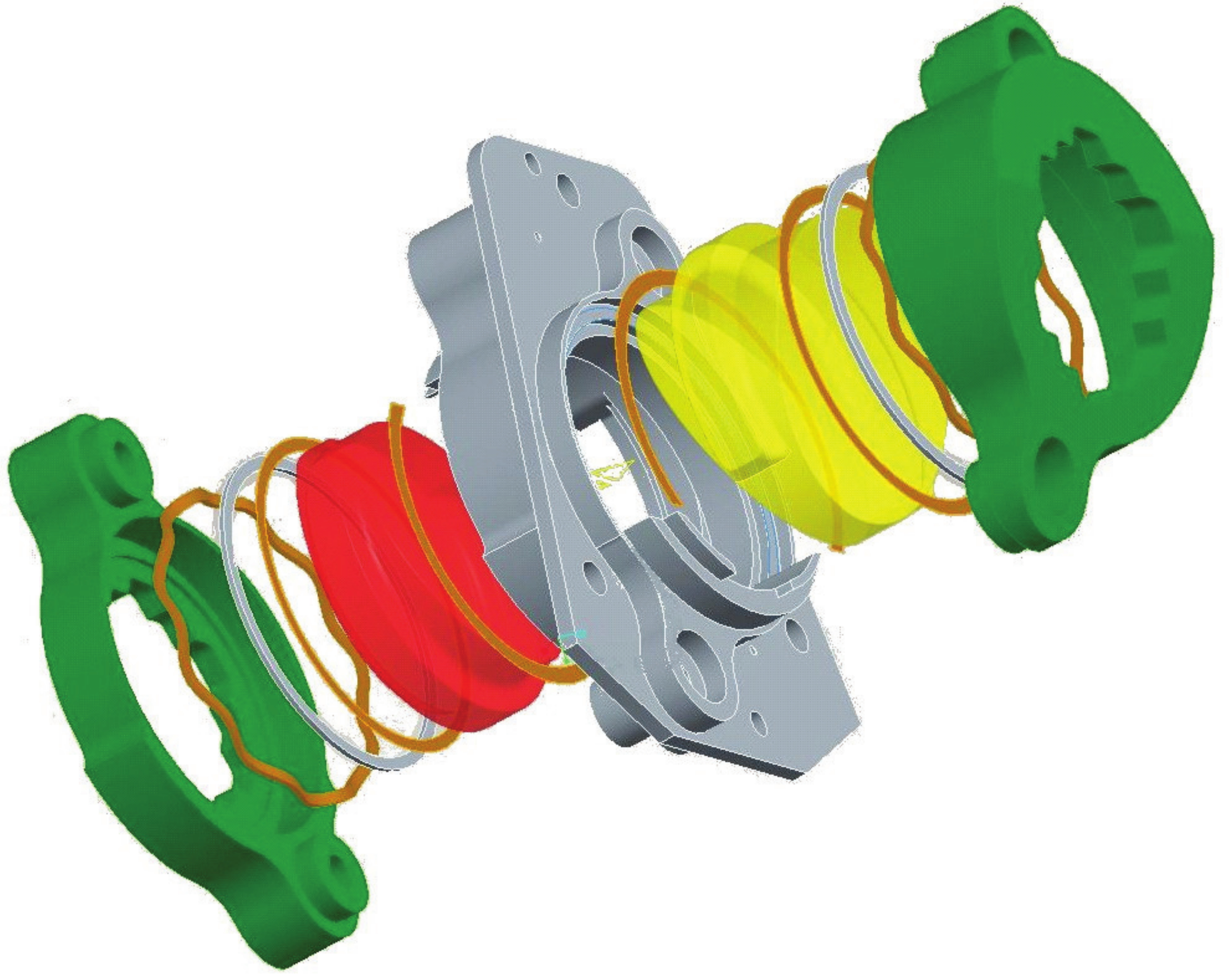}
\caption{Exploded model of the double prism assembly (DPA), showing the prisms in their opto-mechanical structure. The DPA is mounted in the imager filter wheel.}\label{fig:dpa}
\end{figure}

The two prisms are housed within the Double Prism Assembly (DPA; see Figure~\ref{fig:dpa}), which is attached to the imager filter wheel. The DPA is designed to protect the prisms from excessive loads. Copper Beryllium undulated springs press the prisms against bearing surfaces in the holder and, to achieve a more uniform force distribution between spring and prism, a 0.5mm ring is placed on top of the flange. Gold foil of a thickness of 0.025mm provides soft contact to the prism crystalline material. The undulated springs fix the prisms against movement in the axial direction and (via static friction) in the lateral directions. The mass of the Ge prism is 18.8g, the mass of the ZnS prism is 24.1g and the whole DPA has a mass of 71.4g.

The rings, covers and the holder are made out of Al6061 T651 and machined from solid.
An aperture representing the footprint of the JWST astronomical beam at prism-level (including 3.8\% pupil shear) is implemented into the design of the cover.
The radial gap between holder and prism is sufficient to compensate for the different thermal expansion of the different materials and to provide the required accuracy in lateral positioning of the prism. Rotational referencing around the generating axis of the prism is provided by the shape of the flange on each prism.

There is no shutter or way to block light going into the slit while imaging nor is there a way to mask the imaging portion of the focal plane when taking a low resolution spectrum. If there are other point sources in the imaging field at the time of an LRS observation, these will appear as slitless spectra on the imaging fields. This raises the possibility of light scattering into the LRS detector region and potentially detrimental effects of saturation in the imager plane; these issues are discussed in more detail in Section~\ref{sec:scattering}. The location of the spectrum of a source through the slit was chosen such that there would be no overlap with the spectrum of any point sources in the Lyot coronagraph field-of-view.

\section{Measured performance}\label{sec:performance}

As part of the Imager module, the LRS was assembled at CEA Saclay and underwent its first system-level testing there. As neither the Engineering and Test Model (ETM) nor the Verification Model (VM) of MIRI included the DPA, the LRS mode was first tested in the final flight hardware. The successive flight test campaigns offered the opportunity to verify the performance of the LRS, as well as to carry out essential measurements for the calibration of the instrument. In this section we detail the results of these measurements as relevant to the scientific operation of the LRS.

The flight test campaign at CEA Saclay (2009-2010) lacked the full MIRI pre-optics assembly and the range of calibration sources offered by the MIRI Telescope Simulator (\citet{belenguer2008} and Paper II), and was thus not fully representative of the final instrument system. In addition, at this stage the MIRI Imager did not include the flight detectors; radiometric measurements were thus not made. Valuable optical measurements and calibration tests were however carried out in this testing phase.

Following integration with the instrument at the Rutherford Appleton Laboratory (RAL) in 2011, the LRS was extensively tested both optically and radiometrically, with the calibration aids provided in the MIRI Telescope Simulator (MTS). The aim of this campaign was to verify the performance requirements of LRS and to perform the measurements needed for the first calibration data products.
The test data yielded the first versions of the LRS relative spectral response functions, detector pixel flat field, wavelength dispersion relation, field of view measurement and optical image quality. Given the demands from the scientific goals for the LRS, particular attention was paid to the optical and radiometric stability of the LRS spectra, and the potential stray light contamination from bright sources coincidentally present in the MIRI imager field of view.

\subsection{Point spread function}\label{sec:psf}

Because the LRS aims to provide near-diffraction-limited image quality over the 5-10~\um~range, dedicated tests were carried out using the MIRI flight hardware and MTS to measure the spatial point spread function (PSF). Knowledge of the PSF is required for the definition of an optimal spectral extraction procedure, which is an essential step of the LRS data processing pipeline.

Initial measurements of the LRS image quality using a 25~\um~pinhole source showed the presence of an artifact in the wing of the PSF, as well as a degree of astigmatism, degrading the apparent image quality. Analysis and optical modelling showed the former to originate in the pinhole mask of the MTS, outside of MIRI. Optical modelling of the intrinsic PSF of the MTS 25~\um~pinhole indicates a contribution of 1.1 px to the observed LRS PSF at 7.7~\um. Correcting the mean measured FWHM of the spectrum at this wavelength (2.8 $\pm$ 0.1 px) for this effect, we expect the in-orbit image quality to be 2.6 $\pm$ 0.1 px at 7.7~\um. Given the MIRIM scale of 0.11\arcsec/px, this is within 10 to 15\% of the 7.7~\um~diffraction limit of a 6.5-m telescope.

To produce a detailed model of the PSF for extraction and calibration purposes, 39 individual spectra of an unresolved point source (using the 25~\um~pinhole source in the MTS) were used to characterise the PSF shape. Average Gaussian fit parameters to the PSF spatial profile (along the slit direction) were calculated for all spectra on a row-by-row basis, ignoring the rows with a low signal-to-noise and clipping outlier values at $>$ 3 $\sigma$. Standard errors on the PSF were calculated using the uncertainties on each individual fit and performing a Monte Carlo analysis.

\subsection{Spectral distortion}\label{sec:specdist}

Spectrograph optics typically distort the final spectral trace of a point source on the detector. Characterisation and correction of this spatial distortion prior to spectral extraction forms an important part of the data processing and analysis procedure. To study this effect for LRS, we determined the shape of 81 point source spectra at 20 different positions in the slit. The precise location of the spatial centroid at each row was identified using Gaussian fits. The distortion profile was then determined by performing a polynomial fit to the centroid locations. The best fit was obtained using a first-order polynomial, i.e. a straight line, at all positions. The slope of the line is very small, with the spatial centroid of the trace shifting by approximately 1 pixel over the full spectrum ($\sim$ 300 rows). 

There is no evidence for variation in the distortion profile with the position of the source along the length of the slit within the uncertainty on the current measurements ($\sim$20\% variation in measured slopes between the spectra). However, analysis of larger datasets, better detector calibration (flatfielding and background subtraction), and improved pointing accuracy will further refine our knowledge of the distortion profile.

In the LRS calibration pipeline, the distortion is corrected as part of the spectral extraction procedure by applying the appropriate rotational coordinate transformation to the spectrum.

\subsection{Spectral resolving power}\label{sec:specres}

The test campaign at CEA Saclay in 2009-2010 included a finely stepped monochromator scan to measure the as-built spectral resolving power R (=$\lambda$/$\Delta \lambda$) near 7.5 \um. After correcting for the intrinsic linewidth of the monochromator, the LRS resolving power was found to be 95.3 $\pm$ 0.6 at 7.5 \um. Measurements over the range 7-8 \um~are shown in Figure~\ref{fig:lrs_disp_res_meas}. These data were not corrected for the increasing PSF size with wavelength, which may affect the measured value of R. While the resolving power was not measured over the full wavelength range, the data available suggest a linear trend close to the as-designed specification.

\subsection{Wavelength calibration}\label{sec:wavecal}

\begin{figure*}[h]
\includegraphics[width=7.0in.]{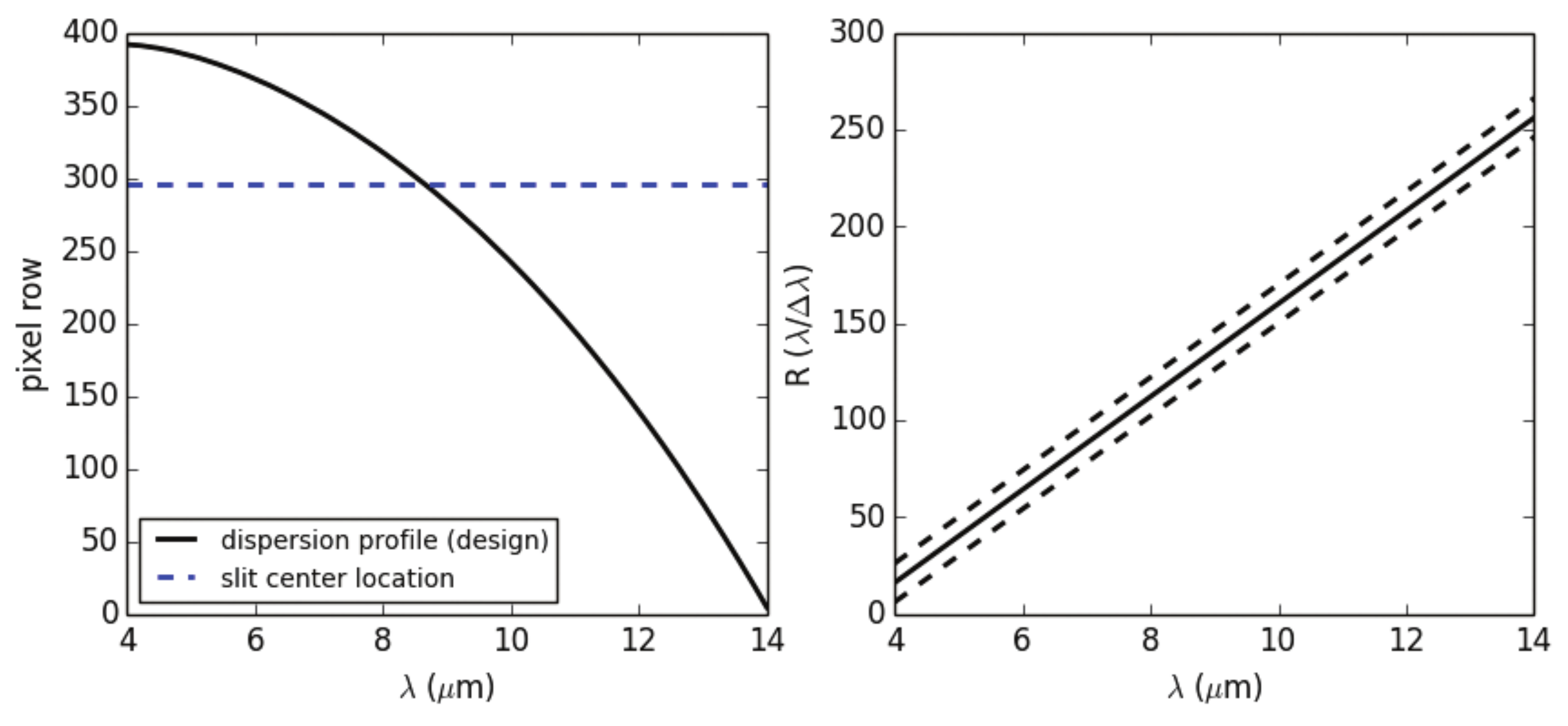}
\caption{Left: As-designed spectral dispersion profile of the LRS between 4 and 14~\micron. The position of the slit center is shown with the dashed line. Right: As-designed spectral resolving power of LRS. The resolving power varies linearly with wavelength, with R=100 at $\lambda$= 7.5~\micron. The dashed lines show the upper and lower bounds of the design specification.}\label{fig:disp_res_design}
\end{figure*}

\begin{figure*}
\includegraphics[width=\textwidth]{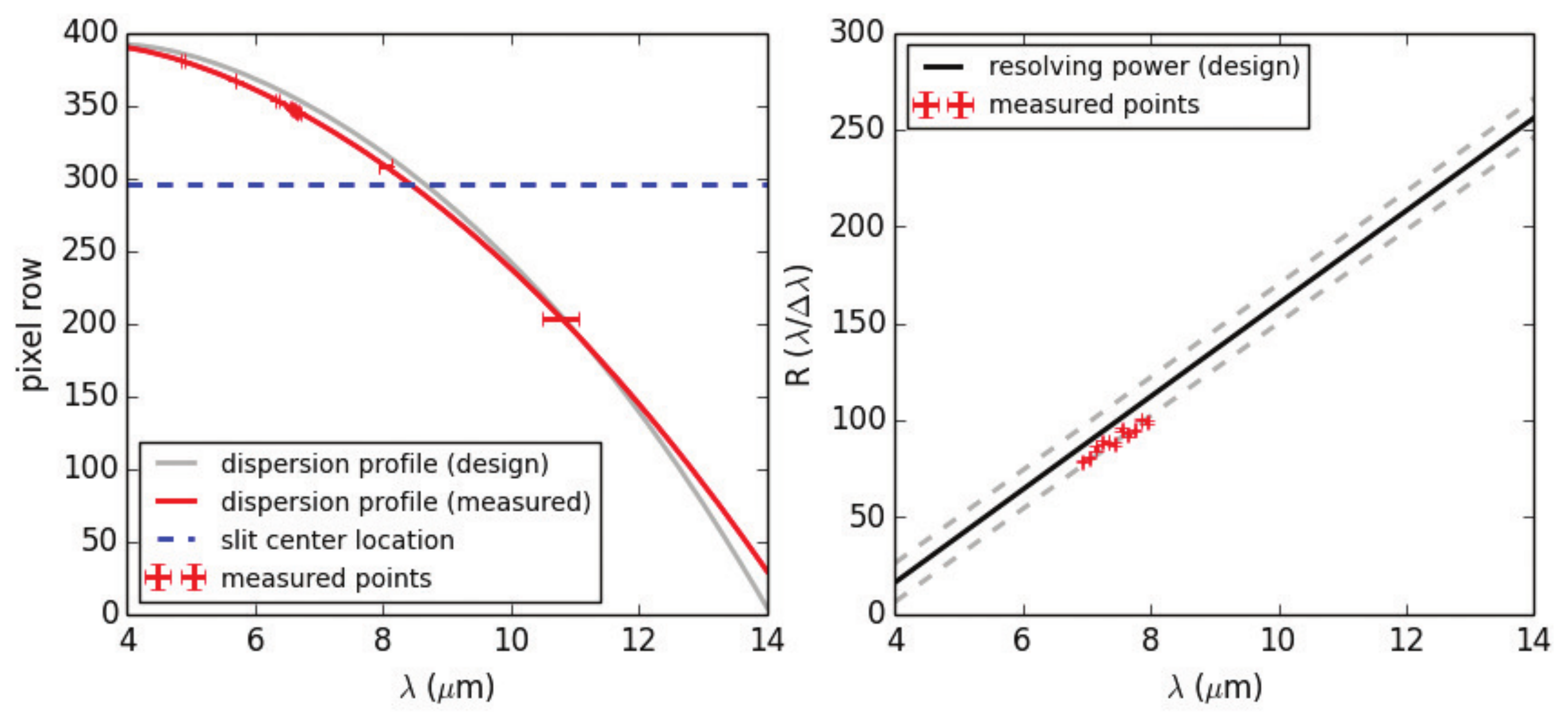}
\caption{Predicted LRS spectral dispersion (left) and spectral resolving power (right), as in Figure~\ref{fig:disp_res_design}. In both plots, points have been added to show the measurements from flight testing. The best-fit spectral dispersion profile is shown in red in the left-hand panel.}\label{fig:lrs_disp_res_meas}
\end{figure*}

A number of measurements were performed during the MIRI test campaigns to determine the spectral dispersion properties of the LRS, required for wavelength calibration of the spectra. The as-designed dispersion of the instrument is strongly non-linear (see Figure~\ref{fig:disp_res_design}), particularly at the shortest wavelengths (4.5 - 6~\um). While the turnover of the dispersion below 4.5~\um~was mitigated by a blocking filter, the steepening of the slope introduces a degeneracy in the pixel-wavelength relation (Figure~\ref{fig:disp_res_design}) near the turnover point. The nominal 5-10~\um~range is dispersed over approximately 135 pixels, giving a mean scaling of $\sim$37~nm/px. Beyond 14.5~\um~the slit spectrum no longer falls onto the detector (or within the subarray in the slitless case); however the instrument throughput effectively limits its performance to $\lesssim$12-13~\um.

Using the available filters in the MTS, the dispersion relation was constrained at a number of points along the LRS range, and compared with the as-designed dispersion. These measurements used spectra of point sources, positioned at the centre of the slit. The differences between the observed and predicted locations of the reference positions were used to fit the dispersion profile to the observations. The final best-fit pixel-wavelength table together with the measurement points is shown in Figure~\ref{fig:lrs_disp_res_meas}. The wavelength calibration relation was found to be accurate to within $<$~10~nm within the 5-10~\um~LRS nominal range. Outside this range the uncertainty rises to 10 to 15 nm longwards of 10~\um, and to $\sim$35~nm for $\lambda$ $<$ 5~\um. The accuracy is currently limited by the lack of unresolved line sources or other strong reference features over the full LRS wavelength range.

Of critical importance for the wavelength calibration is the precise knowledge of the source position in the slit. During testing this knowledge was limited by the calibration of the point source scanning mechanism coordinate system with respect to the detector, to approximately 0.2~px. 

During the mission, the LRS wavelength calibration will benefit from the excellent pointing accuracy of JWST and the availability of astronomical targets with unresolved emission lines in the LRS wavelength range. Based on current data, for the expected pointing accuracy of 4.6 mas, we estimate an accuracy of the in-flight wavelength calibration of 1-2 nm within the nominal wavelength range.
Further measurements over the course of the test campaigns showed the dispersion to have excellent stability over time, varying by no more than 1.7 nm over any given 7-day period. There was no evidence of any changes in dispersion profile with source position along the slit. However, when the source is moved away from the slit centre in the across-slit direction the PSF quickly becomes truncated, affecting the accuracy of the wavelength calibration. Spectra taken of a point source placed at $\pm$1.25~\um~from the slit centre showed a deviation of $\sim$30 nm in dispersion at a given reference position (2.5~\um~ corresponds to the expected pointing accuracy of 4.6 mas). Slit spectroscopy with MIRI therefore requires careful positioning of the source in the centre of the slit, and detailed knowledge of the telescope pointing.

\subsection{Flatfielding}\label{sec:ff}

Flat fields will be applied to LRS observations to remove variations in both the pixel-to-pixel and low spatial frequency response of the detector.  The pixel-to-pixel variations will be measured using the calibration source in the Interface Optics and Calibration IOC unit (Paper II). Since this source does not uniformly illuminate the focal plane, low spatial frequency variations in the pixel flats will be removed by normalizing the flat using a 2-dimensional low-order polynomial. At LRS wavelengths, the astronomical background is expected to dominate the telescope thermal emission~(Paper IX, \citet{paper8b}); therefore, low spatial frequency variations will be measured using observations of the sky to produce sky flats. Large numbers of 2D spectra will be combined using outlier rejection to eliminate sources and cosmic ray artefacts. This 'super-flat' may have very high signal-to-noise depending on the number of fields that are co-added. The effectiveness of the super-flat will depend on stray light control in the telescope and will be evaluated on orbit.

In the test campaigns, the flatfield was produced by imaging the calibration source through broad- and narrow-band filters to illuminate the LRS spectral region on the detector. The low spatial frequency variation was fitted using a low-order polynomial on a row-by-row basis to correct for the non-uniform illumination pattern of the source, producing a separate calibration image. Following this correction for large scale variation, weighted means of flatfields derived from independent exposures were taken to produce the pixel-to-pixel flatfield, with weights determined by the errors of the individual flatfields. The method was tested with 25 exposures on a small detector region and using a single broad-band filter, and resulted in a residual standard error of $\sim$0.5 $\%$.

\subsection{Relative Spectral Response Function}\label{sec:RSRF}

Relative Spectral Response Functions (RSRFs) will be applied to LRS observations to spectrophotometrically calibrate the data. The absolute calibration of the MIRI (and the other JWST instruments) will be based on astronomical observations of very hot (white dwarfs and OB stars), hot (A-type), and warm (G-type (solar-like) stars) to quantify differences between stellar and model spectra and to control for systematic uncertainties in stellar atmosphere modeling. All of the proposed JWST calibrators have been observed using the Hubble and Spitzer Space Telescopes with high signal-to-noise (S/N $>$ 100). For a subset of these calibrators, the existing Hubble and Spitzer calibrations are consistent to within 2\% \citep{bohlin11}. The predicted spectra for these calibrators are drawn from the CALSPEC database\footnote{http://www.stsci.edu/hst/observatory/cdbs/calspec.html} using models for White Dwarfs \citep{bohlin95} and measurements for A- and G-type stars \citep{rieke08} at wavelengths shorter than 2.5~\um~and models at longer wavelengths. From an intercomparison of the measurements, it will be possible to test and refine the RSRFs.

Although the optimal RSRFs can only be determined in orbit using astronomical standards, on-ground RSRFs have been determined using the MTS. This simulator incorporates a black body source with adjustable temperature to produce different flux levels, as well as provision for both point and extended sources to be fed to the instrument. During the MIRI test campaign, LRS point source measurements were taken at different black body temperatures between 300~K and 800~K using a 100~$\mu$m diameter pinhole (1.428 pixels on the detector). As expected, the temperature dependence of the RSRF is negligible. After background subtraction, the LRS spectra were extracted from these slope images using aperture photometry, and the wavelength calibration applied (see Section~\ref{sec:wavecal}). Both slit and slitless spectroscopy modes were tested.

\begin{figure}[h]
%\centering\includegraphics[width=5.0in.]{RSRF_800K_PASP.png} 
%\centering\includegraphics[width=5.0in.]{FIG6_RSRF_800K_PASP.png} 
\centering\includegraphics[width=5.0in.]{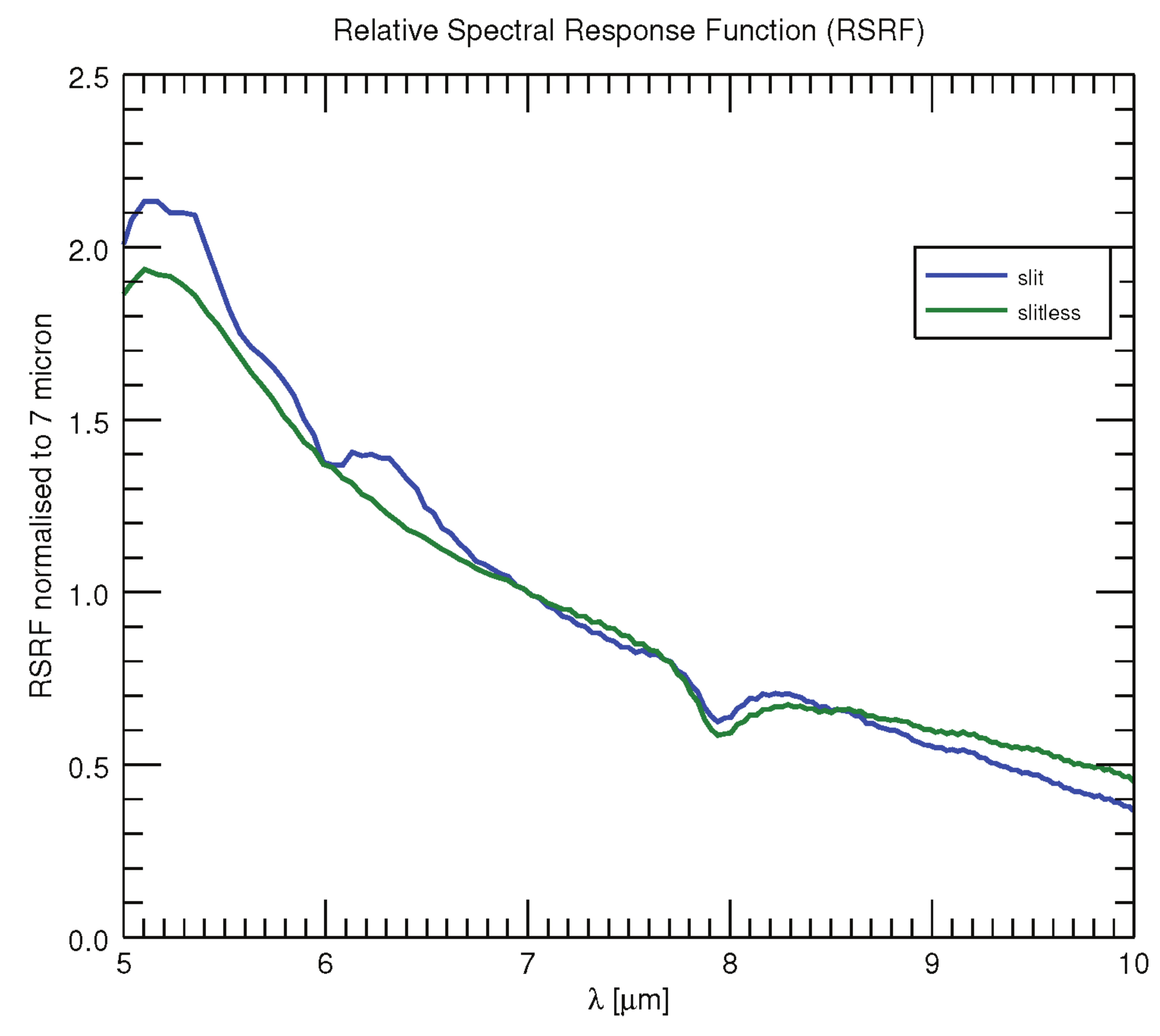} 
\caption[RSRF from flight model campaign LRS measurements.]
{Relative Spectral Response Function (RSRF) for the LRS. 
The RSRF is the spectral response function (SRF) normalised to the value at 7~$\mu$m. Data for slit (blue) and slitless (green) spectroscopy are shown.}
\label{Fig:RSRF_Tmean_PASP}
\end{figure}

\begin{figure}[h]
%\centering\includegraphics[width=5.0in.]{PCE_800K_PASP.png} 
%\centering\includegraphics[width=\textwidth]{FIG9_PCE_800K_PASP.png} 
\centering\includegraphics[width=\textwidth]{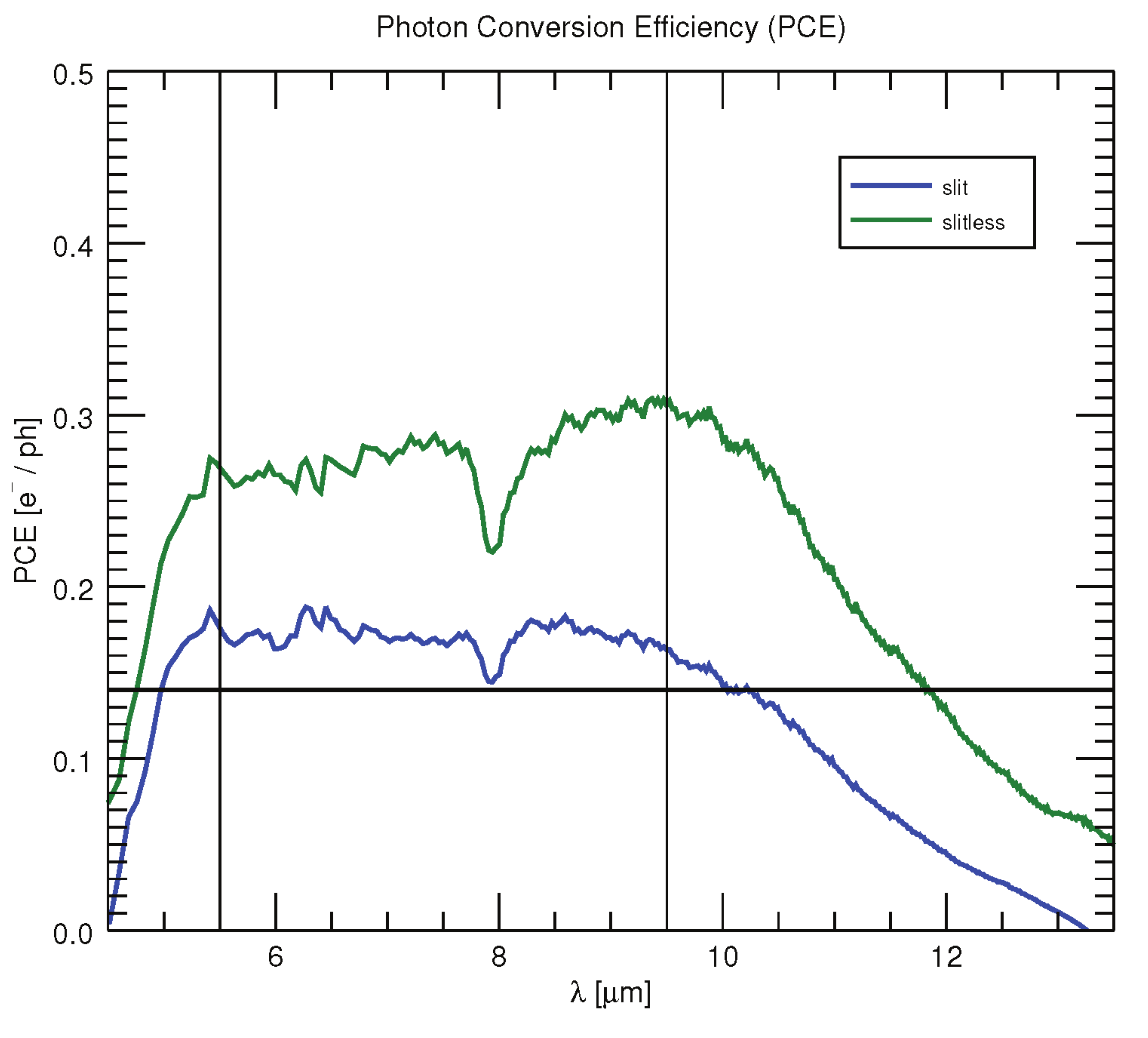} 
\caption[PCEs from FM LRS measured point source fluxes.]
{Photon conversion efficiencies (PCE) from LRS measured point source fluxes during flight testing. The blue curve represents slit spectroscopy, the green curve slitless.
The two vertical and horizontal black lines mark the range over which the MIRI PCE requirement was specified, i.e., $>$ 0.14 between 5.5 and 9.5~\micron.}
\label{Fig:PCE_Tmean_PASP}
\end{figure}

Figure~\ref{Fig:RSRF_Tmean_PASP} shows the RSRF for the 800~K measurement, representing the Spectral Response Function (SRF) normalized to 7~$\mu$m. The SRF is the ratio between the measured point source flux and the flux reaching the MIRI imager entrance focal plane. The latter is calculated with MTSSim, a software tool used to simulate the flux output of the MTS. The blue line in Figure~\ref{Fig:RSRF_Tmean_PASP} shows the RSRF for slit spectroscopy, and the green line shows the slitless spectroscopy measurement. The SRF values at 7~$~\mu$m are approximately $5.5 \times 10^4$~(DN/s)/Jy for slit and $9 \times 10^4$~(DN/s)/Jy where 1~DN = 5.5~e$^-$  (Paper VIII). As the pinhole used for these measurements was not a perfect point source, the flux losses within slit measurements were larger than expected for a point source. Our slit measurements are therefore not representative for in-orbit point source slit spectroscopy. To get a proper SRF estimate for LRS slit observations we used the slitless measurements, corrected those using the LRS slit mask transmission profile and applied the expected slit throughput for a perfect JWST PSF (Paper IX). Note that the drop in RSRF around 8~$\mu$m is due to a discrepancy between the MTS model and true output; this will not be present in the in-orbit SRF. In addition, an empirical correction factor of 1/0.55 was applied to properly represent the real MTS flux output (Paper IX). This factor has been taken into account for both the RSRF and Photon Conversion Efficiency (see Figure~\ref{Fig:PCE_Tmean_PASP}) calculations.

\subsection{Scattered light from bright sources in imager plane}\label{sec:scattering}

When observing with LRS the imager portion of the focal plane remains open, producing dispersed images of any sources serendipitously located in this part of the field of view. Light can thus be scattered from these sources onto the LRS part of the detector; dedicated measurements were carried out during the flight test campaign to estimate the potential impact of this on LRS flux measurements. These measurements, in which the focal plane was illuminated with the MTS extended blackbody source at temperatures of 300 and 400 K, indicate that $<$0.1\% of the illumination of the imager part of the detector is scattered into the region of the LRS spectrum. 

These tests however revealed another important issue: although low temperature levels had been chosen (300~K and 400~K) for the extended source illumination, the imager part of the detector almost completely saturated. This affected the measured slopes in the LRS spectra, which made the extraction of reliable LRS spectra impossible. It is therefore important to avoid saturation in any part of the detector when observing with the LRS.

To quantify this flux limit, we derived conversion factors from DN/s/px to Jy/arcsec$^2$ for the extended source data. In principle, these conversion factors are the reciprocal of the point source Spectral Response Function (SRF) divided by the pixel area on sky (0.11~arcsec$^2$) and could thus be compared with the SRF values derived in Section~\ref{sec:RSRF}
For low resolution spectroscopy with the double prism, the maximum flux of an extended source in the imager field near the LRS strip is 140 mJy/arcsec$^2$ for a 300 K black body at 7.7~$\mu$m assuming 2 frames per integration and full array readout.

\subsection{Fringing}\label{sec:fringing}

Spectral fringes as a result of optical interference are a common characteristic of infrared spectrometers~\citep{lahuis2003, kester2003}. The primary source is usually the detector itself, where the semiconductor layer with an optical thickness of a few $\mu$m is a very efficient resonator at infrared wavelengths. Within the MIRI, fringes are observed originating in the detector surface and filters in the MIRI optical path. In the medium resolution spectrometer (Wells et al. (2014): Paper VI), the fringes are resolved and have a medium-to-high modulation depth. However, in the LRS the fringes are unresolved and blend in with the SRF. However, the optical properties of the LRS detector are similar to those of the MRS short wavelength detector. We therefore use the MRS fringe characteristics to assess the impact of the fringes on LRS spectra. 

\begin{figure}[h]
%\centering\includegraphics[width=5.0in.]{LRS_delta_fringes.png} 
\centering\includegraphics[width=7.0in.]{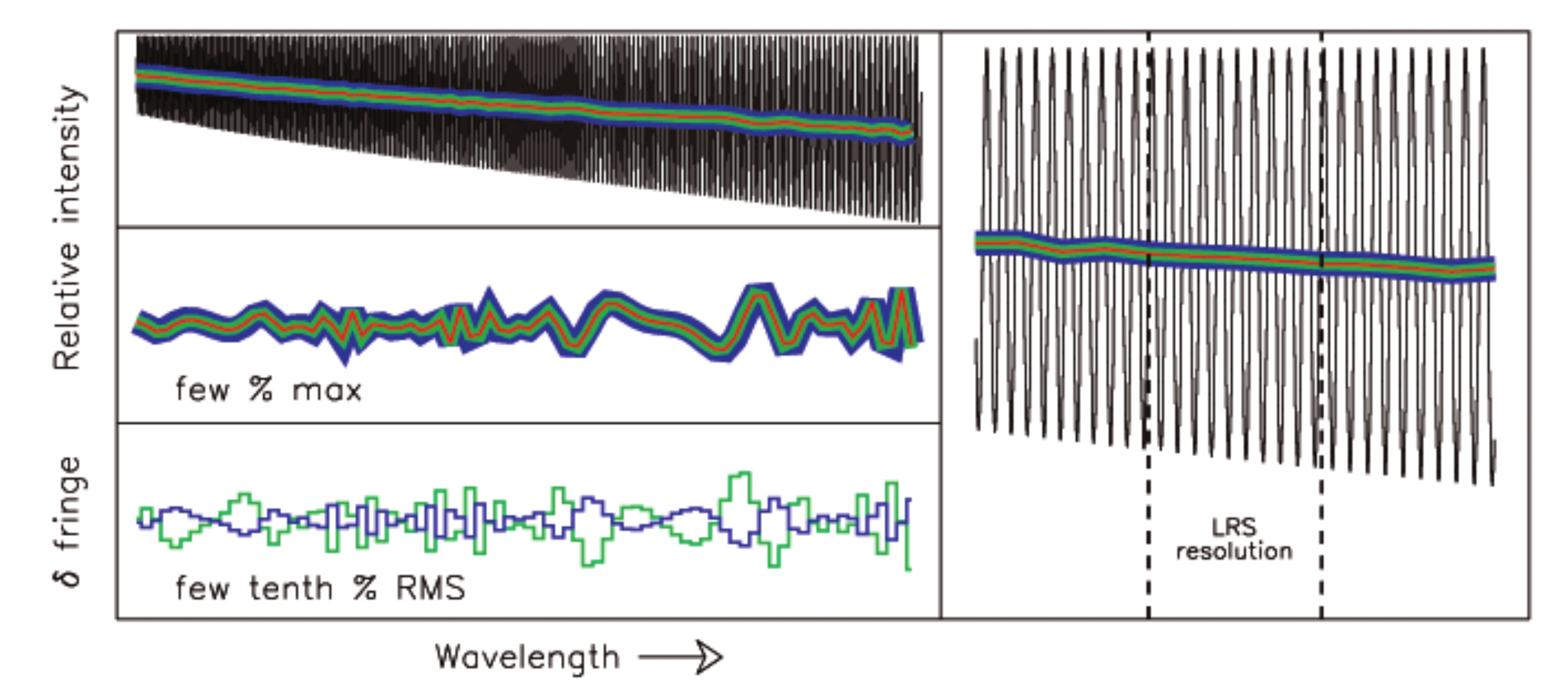} 
\caption[Synthesized LRS fringes over a fraction of the LRS band.]
{Synthesized LRS fringes over a fraction of the LRS band. Th panel on the right shows a strong zoom-in down to the LRS resolution. Black shows the resolved fringe pattern (based on 
Integrated Field Unit MRS Short Wave measurements) and the red, green and blue curves show the fringes at LRS resolution for three pointings (red at nominal pointing and blue and green with small sub-pixel offsets). The lower left panel shows the variation in the fringe patterns as a result of small pointing offsets.}
\label{fig:fringes}
\end{figure}

Figure~\ref{fig:fringes} shows how the fringes may appear as an SRF modulation of $<$ 1\% at the shortest to a maximum of a few percent at the longest wavelengths. This modulation is the result of a varying ratio of the LRS resolution element to the fringe width. Small pointing offsets (even a few \% of a pixel) result in a small shift in the wavelength and thus the observed pixel fringe pattern. The amplitude of this variation is typically of order 10$^{-4}$. For nominal observations, pointing-induced variations will therefore not be an issue and the fringe pattern will directly be divided out with the SRF. However for extreme S/N observations (e.g. exoplanet spectroscopy) pointing jitter or drifts can become a significant noise component if not properly corrected. Therefore the development of a modeling framework complemented with targeted in-orbit calibration observations will be important.

\section{LRS operations}\label{sec:ops}

The operational concept of the MIRI, including calibration and data reduction strategies, is described in detail in Paper X. In this section we focus on two LRS-specific operational issues, target acquisition and the slitless operation of the spectrometer.

\subsection{Target acquisition}\label{sec:target}

Acquiring a point source into the LRS slit centre will make use of a small target acquisition region in the imager portion of the detector located $<$ 20\arcsec~from the slit centre (within the maximum distance that can be covered by a small angle maneuver (SAM) with the telescope, for which no new guide stars are required). During target acquisition, the source is placed within this region, and imaged with a suitable imaging filter. A floating window centroiding algorithm is then employed to determine the exact position of the source within the acquisition region.

During MIRI flight testing a 91 $\times$ 91 px (10 $\times$ 10\arcsec) region was defined in the imager plane, lying fully within a 20\arcsec~radius from the LRS slit centre. The approximate location of this region is shown in Figure~\ref{Fig:detector_with_spectrum}. From the acquisition images the centroiding algorithm yielded a centroid accuracy of 0.1 and 0.4 px for bright and faint sources, respectively, allowing in the first instance for the source to be positioned at the slit centre to $\leq$ 0.2 px. 

The target acquisition procedure is further described in Paper X.

\subsection{Slitless spectroscopy with LRS}\label{sec:slitless}

\begin{figure}[h]
%\centering\includegraphics[width=5.0in.]{mirim_lrs.png} 
\centering\includegraphics[width=7.0in.]{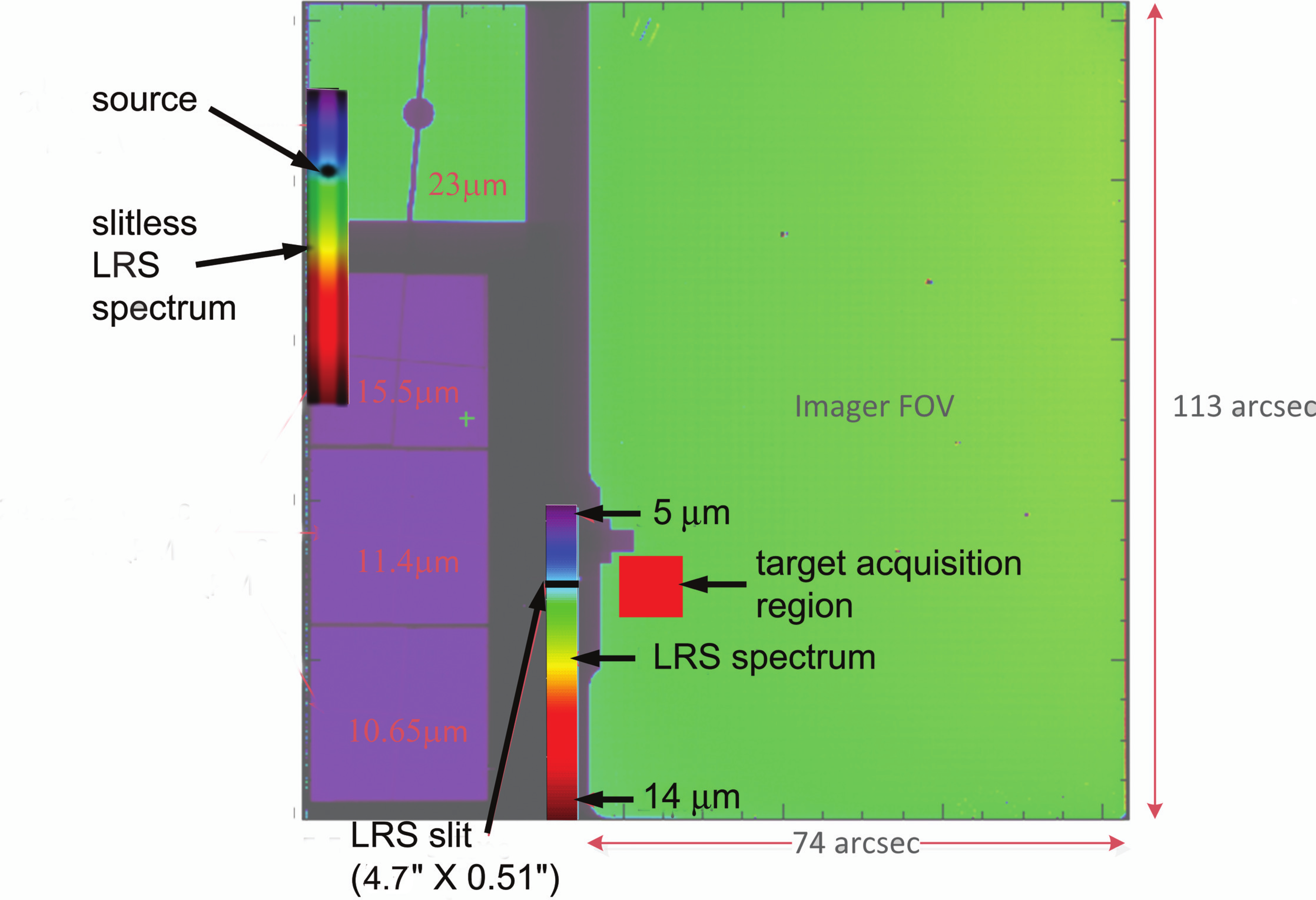} 
\caption{Schematic view of the MIRIM focal plane highlighting the locations of the (fixed) 4\farcs7  LRS slit, and the LRS 7$''$ wide slitless readout window. The slit centre position is located in pixel (321, 296); the lower left-hand corner of the slitless subarray is located in (1, 529). The direction of dispersion is indicated at both locations. The red box shows the approximate location of the 10\arcsec $\times$ 10\arcsec target acquisition region for slit spectroscopy (see Section~\ref{sec:target}).}\label{Fig:detector_with_spectrum}
\end{figure}

This section will describe the rationale for inclusion of the slitless spectroscopy mode, and the operational concept where it differs from slit spectroscopy. 
As the LRS double prism is integrated into the MIRI imager filter wheel, low resolution spectra can be 
obtained on any source within the imager field-of-view, without using the LRS slit. Currently, only one operational mode is foreseen for slitless spectroscopy with the LRS, where a source is placed in a dedicated LRS slitless detector region, shown in
Figure~\ref{Fig:detector_with_spectrum}. In contrast to the slit spectroscopy mode, this region is read out as a subarray for slitless spectroscopy.

The main scientific driver for slitless spectroscopy with the LRS is high 
precision spectro-photometry of bright nearby stars with transiting planets to 
obtain spectra of exoplanet atmospheres. The contrast between exoplanet and host 
star of the transiting systems that will be observed with MIRI can be as low as 
10$^{-4}$. This contrast ratio needs to be reached to detect the 
exoplanetary atmospheric signal, but can be achieved over a number of 
transits/occultations. The intrinsic stability of the LRS should be at least as 
good or better than the exoplanet-to-host star contrast ratio over an interval 
equivalent to the time taken to observe the transit (typically a few hours). 

A typical problem with spectroscopic observations using a narrow (relative to the 
PSF) slit is throughput variations due to telescope pointing uncertainties and 
drifts. This effect is one of the dominant sources of systematic noise in the 
low-resolution spectra taken with Spitzer of transiting exoplanets  
\citep[e.g.,][]{swain2008}. For Spitzer this resulted in a limitation of the maximum S/N 
ratio of $\sim$200, and only by correcting the throughput variations could higher 
 S/N ratios be achieved ($\sim$2000). As the LRS slit width compared to 
the width of the PSF is similar as that for the low-resolution spectrometer of 
Spitzer in the 5 to 12~$\mu$m wavelength range, similar problems are expected. The natural solution, therefore, to reach the required 
spectro-photometric precision is to operate the LRS in a slitless mode, thus 
preventing any pointing induced throughput variations. 

A second problem is that the required S/N ratios can only be achieved for relatively nearby, bright transiting systems. The detector readout time used for the LRS slit observations  is 2.775~sec. Assuming a minimum of 2 up-the-ramp samples and that the brightest pixel will not exceed 60~\% of the full storage capacity, the saturation limit at 5.5~$\mu$m is estimated to be 63~mJy (see further Paper IX, section 3). Assuming an M0 star, this would imply a limiting magniture of K$ = 8.3$. This would mean that many of the currently known exoplanet host stars are too 
bright to be observed with standard LRS slit observations. This situation is 
likely to become more challenging with the TESS space mission~\citep{Ricker2010}, expected to discover many bright nearby transiting systems, which would be ideal targets for 
spectroscopic investigations. To address this issue a subarray was defined on the MIRI imager focal plane array, substantially reducing the readout time for exposures.
The minimum readout  time is set by the number of detector columns to be read out.

The LRS slitless subarray, as shown in Figure~\ref{Fig:detector_with_spectrum}, is the optimal solution between the requirement for fast readout times and a large enough detector area to get a  proper spectral image. The readout time of the LRS slitless subarray is 0.159~sec (Paper VIII). Taking into account the assumed $\sim$30\% flux losses due to the slit in the calculation of the bright source limits (see  Paper IX) and again, assuming a minimum of 2 up-the-ramp samples, the saturation limit at 5.5~\um~would be around 770 mJy, which results in a limiting magnitude of K= 5.65 assuming an M0 star. This limiting magnitude would mean that almost all known transiting systems could be observed.

\subsection{Slit versus Slitless Spectroscopy Operation}\label{sec:slitvslitless}

\begin{figure}[h]
%\centering\includegraphics[width=5.0in.]{LRS_transmission_slit_PASP.png} 
\centering\includegraphics[width=5.0in.]{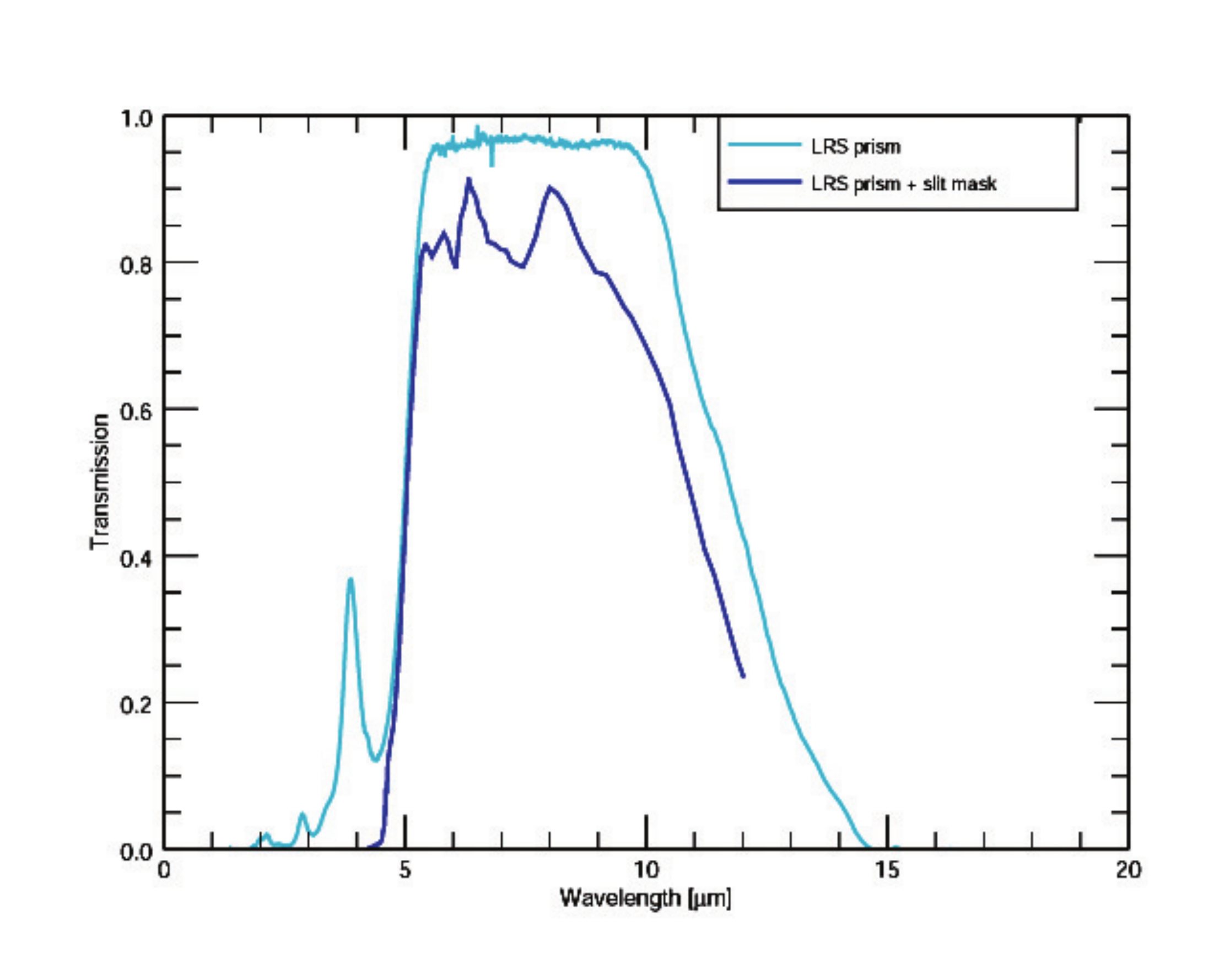} 
\caption[Transmission curves for LRS prism and LRS prism plus slit.]{Measured transmission curves for the LRS double prism (cyan), and for the LRS double prism + slit (blue). The blue curve was only measured up to 12 micron.}\label{Fig:LRS_trans_slit}
\end{figure}

Figure~\ref{Fig:LRS_trans_slit} shows the combined transmission of the LRS double prism and the slit mask. It shows the measured transmission curves of the LRS prism (slitless spectroscopy); 
the combined transmission curve of LRS prism and slit mask (slit spectroscopy); and the ratio of both curves, which gives the 'slit mask only' transmission. This illustrates the expected light loss from the presence of the slit mask as compared with slitless spectroscopy, particularly at longer wavelengths ($\lambda$ $>$ 9 \micron). At 12~\um, the transmission for slitless spectroscopy is approximately 60\% higher than in the slit + slit mask case.

Subarray-specific reference files will be obtained for and applied to the slitless subarray data in the JWST pipeline, including darks and flat fields. Since slitless mode has a similar sensitivity as slit mode, the majority of the spectrophotometric calibrators selected for LRS with the slit will also be appropriate for LRS operated slitless.

The Photon Conversion Efficiency (PCE) is another difference between the modes. The PCE is the ratio between the electrons measured at the detector relative to the number of photons reaching the MIRI imager entrance focal plane.
Figure~\ref{Fig:PCE_Tmean_PASP} shows the PCE for slit and slitless spectroscopy, derived using the same flight test data and analysis methods as for the RSRF calculation (see Section~\ref{sec:RSRF}); this includes all MIRI optics and the flight model detector features. As for the RSRF, the PCE is not temperature dependent and the feature around 8~\micron~is only of artificial nature, i.e. will not be present in flight. The absence of a slit leads to a higher PCE for slitless operation. First, some photons are lost due to the 
transmission features of the slit mask in the MIRIM entrance focal plane (Figure~\ref{Fig:LRS_trans_slit}). In addition flux may be lost in cases where the source is not perfectly centred in the slit and/or the source is marginally extended (as was the case with the test point source in the MTS).

\section{Summary}

The MIRI Low Resolution Spectrometer provides near-diffraction-limited spectroscopy at R$\sim$100 over a nominal wavelength range of 5-10~\um~but with useful throughput to 12~\um. The module can be operated in slit or slitless mode, with the latter particularly suitable for exoplanet transit observations. The slit measures 0.47 $\times$ 5.1\arcsec~on sky. Optical calibration of LRS spectra requires detailed knowledge of the source position in the slit; to this end a target acquisition region in the imager plane has been defined from where a well-centroided source can be accurately placed at the slit centre via a small angle manoeuver with the telescope.

Dedicated measurements indicate the on-board calibration source can be used to obtain accurate flatfields, correcting the pixel-to-pixel gain variations to within 0.5\%. The LRS photon conversion efficiency was measured to be $>$ 14\% , the requirement, but  $\leq$0.3 from 5-10~\um. Fringing, which is known to occur in the MIRI detectors, will not significantly affect the observations. The operation of LRS in slitless mode offers advantages over use of the slit for some types of observation. All calibration measurements were performed for both modes of operation. An operational concept has been developed to optimise slitless operation of the LRS, including designating a dedicated detector subarray to enable slitless spectroscopy of bright targets.

The MIRI test campaigns have provided reliable measurements of the LRS performance, and indicated that this MIRI module is expected to perform as designed.

\section{Acknowledgments}
The work presented is the effort of the entire MIRI team and the enthusiasm within the MIRI partnership is a significant factor in its success. MIRI draws on the scientific and technical expertise many organizations, as summarized in Papers I and II. 

We would like to thank the following National and International
Funding Agencies for their support of the MIRI development: NASA; ESA;
Belgian Science Policy Office; Centre Nationale D'Etudes Spatiales;
Danish National Space Centre; Deutsches Zentrum fur Luft-und Raumfahrt
(DLR); Enterprise Ireland; Ministerio De Economi{\'a} y Competividad;
Netherlands Research School for Astronomy (NOVA); Netherlands
Organisation for Scientific Research (NWO); Science and Technology Facilities
Council; Swiss Space Office; Swedish National Space Board; UK Space
Agency.

\clearpage

\clearpage

\clearpage

\clearpage

\clearpage

\clearpage

\clearpage

\clearpage

\clearpage

\begin{thebibliography}{}
%\bibliography{references}

\bibitem[Baffa et al. (2001)]{baffa2001}
Baffa, C. et al. 2001, A\&A, 378, 722

\bibitem[Belenguer et al. (2008)]{belenguer2008}
Belenguer, T. et al. 2008, SPIE, 7010, 95

\bibitem[Bohlin (1995)]{bohlin95}
{Bohlin}, R.~C. 1995, AJ, 110, 1316

\bibitem[{{Bohlin} {et~al.}(2011){Bohlin}, {Gordon}, {Rieke}, {Ardila},
  {Carey}, {Deustua}, {Engelbracht}, {Ferguson}, {Flanagan}, {Kalirai},
  {Meixner}, {Noriega-Crespo}, {Su}, \& {Tremblay}}]{bohlin11}
{Bohlin}, R.~C., {Gordon}, K.~D., {Rieke}, G.~H., {Ardila}, D., et al. 2011, \aj, 141, 173

\bibitem[{{Bouchet} {et~al.}(2014)}]{paper3}
{Bouchet}, P., {et~al.} 2014, PASP, this volume, {\bf Paper III}

\bibitem[Donati (1862)]{donati1862}
Donati, C. B. 1862, Il Nuovo Cimento, 15, 292

\bibitem[Fischer et al. (2008)]{fischer2008}
Fischer, S. et al. 2008, SPIE, 7010, 103

\bibitem[{{Glasse} {et~al.}(2014)}]{paper8b}
{Glasse}, A.~C.~H., {et~al.} 2014, PASP, this volume, {\bf Paper IX}

\bibitem[{{Gordon} {et~al.}(2014)}]{paper9}
{Gordon}, K., {et~al.} 2014, PASP, this volume, {\bf Paper X}

\bibitem[{{Kester}(2003)}]{kester2003}
{Kester}, D.~J.~M. 2003, in ESA Special Publication, Vol. 481, The Calibration
  Legacy of the ISO Mission, ed. L.~{Metcalfe}, A.~{Salama}, S.~B. {Peschke},
  \& M.~F. {Kessler}, 243

\bibitem[{{Lahuis} \& {Boogert}(2003)}]{lahuis2003}
{Lahuis}, F., \& {Boogert}, A. 2003, in SFChem 2002: Chemistry as a Diagnostic
  of Star Formation, ed. C.~L. {Curry} \& M.~{Fich}, 335

\bibitem[{{Ressler} {et~al.}(2014)}]{paper7a}
{Ressler}, M., {et~al.} 2014, PASP, this volume, {\bf Paper VIII}

\bibitem[{{Ricker} {et~al.}(2010){Ricker}, {Latham}, {Vanderspek}, {Ennico},
  {Bakos}, {Brown}, {Burgasser}, {Charbonneau}, {Clampin}, {Deming}, {Doty},
  {Dunham}, {Elliot}, {Holman}, {Ida}, {Jenkins}, {Jernigan}, {Kawai},
  {Laughlin}, {Lissauer}, {Martel}, {Sasselov}, {Schingler}, {Seager},
  {Torres}, {Udry}, {Villasenor}, {Winn}, \& {Worden}}]{Ricker2010}
{Ricker}, G.~R., {Latham}, D.~W., {Vanderspek}, R.~K., {Ennico}, K.~A.,
 et al. 2010, B. A A. S., 42, American Astronomical Society
  Meeting Abstracts \#215, \#450.06

\bibitem[Rieke et al. (2008)]{rieke08}
Rieke, G.~H., et al. 2008, AJ, 135, 2245

\bibitem[Rieke et al. (2014)]{rieke2014}
Rieke, G. H. et al. 2014, PASP, this volume, {\bf Paper I}

\bibitem[Rossi et al. (2008)]{rossi2008}
Rossi, L. et al. 2008, SPIE, 7018, 23

\bibitem[Swain et al. (2008)]{swain2008}
Swain, M.~R., Bouwman, J.,  Akeson, R.~L., Lawler, S., \& Beichman, C. A. 2008, ApJ, 674, 482

\bibitem[Wells et al. (2014)]{wells2014}
Wells, Martyn et al. 2014, PASP, this volume, {\bf Paper VI}


\end{thebibliography}
\end{document}